# Comments on "New Insights from the 2003 Halloween Storm into the Colaba 1600 nT Magnetic Depression during the 1859 Carrington Storm" by S. Ohtani (2022)

Bruce T. Tsurutani[1,*], Gurbax S. Lakhina[2], and Rajkumar Hajra[3]

[1]Retired, Pasadena, CA, USA.
[2]Indian Institute of Geomagnetism, Navi Mumbai, India.
[3]Indian Institute of Technology Indore, Indore, India.

[*]Corresponding author: Bruce Tsurutani (bruce.tsurutani@gmail.com)

**Key Points:**

- We disagree with Otani (2022) that an interplanetary sheath magnetic field could have caused the Carrington storm spike at Colaba.
- We argue that the sheath fields would be too low to have created the Carrington magnetic storm.
- There is no evidence that interplanetary sheath magnetic fields can be amplified by up to a factor of 10x.

**Abstract**

The Colaba, India ~-1600 nT magnetic spike caused by an interplanetary sheath magnetic field inducing a "dayside R1-field aligned current wedge" during the Carrington magnetic storm proposed by Ohtani (2022, https://doi.org/10.1029/2022JA030596) seems highly improbable. Normal interplanetary magnetic field intensities of ~5 nT have previously been shown to be sufficient to explain the ~+120 nT $SI^+$ observed at Colaba during the storm (Tsurutani et al., 2018, https://doi.org/10.1002/2017JA024779). Magnetohydrodynamic theory (Kennel et al., 1985, https://doi.org/10.1029/GM034p0001) predicts a maximum of 4x magnetic field compression by a fast shock, giving an interplanetary sheath field of ~20 nT, a value too low to support the Ohtani (2022) hypothesis. The Ohtani (2022) (and Siscoe et al. 2006, https://doi.org/10.1016/j.asr.2005.02.102) claim of a further 10x amplification of the interplanetary sheath fields has not been verified in near-Earth interplanetary sheaths. The original (Tsurutani et al., 2003, https://doi.org/10.1029/2002JA009504) hypothesis that an ICME magnetic cloud having southward magnetic fields of ~90 nT caused the Carrington magnetic storm main phase of peak SYM-H/Dst = -1760 nT seems more likely. The short time between the $SI^+$ and the storm main phase onset implies a foreshortened interplanetary sheath. The extremely rapid recovery of the magnetic storm was hypothesized by Tsurutani et al. (2018, https://doi.org/10.1002/2017JA024779) as being due to nonlinear ring current losses. We point out that the Hydro-Quebec 1989 storm was caused by multiple shock-sheaths and magnetic clouds (Lakhina & Tsurutani, 2016, https://doi.org/10.1186/s40562-016-0037-4) unlike the interplanetary causes of the Carrington storm. The Hydro-Quebec event was a "stealth" magnetic storm.

**Plain Language Summary**

Tsurutani et al. (2018) have previously shown that an ordinary interplanetary plasma density of ~5 $cm^{-3}$ and magnetic field of ~5 nT are consistent with the observed Colaba $SI^+$ of ~+120 nT. A maximum possible 4x shock compression would give a sheath field strength of ~20 nT, a value too low to have caused the Carrington magnetic storm and the Colaba magnetic decrease of 1600 nT as suggested by Ohtani (2022). The cause of the Carrington storm was most probably a Bz ~ -90 nT component inside an interplanetary magnetic cloud (MC) (Tsurutani et al., 2003; Lakhina et al. 2012). The Siscoe et al. (2006) hypothesis of further sheath magnetic field amplification by a factor of up to ~10x has not been observed (no significant amplification has been noted at all). It is concluded that the SYM-H/Dst value of the Carrington storm was -1760 nT, the original value determined in Tsurutani et al. (2003).

**Comment**

We commend the scholarly work by S. Ohtani for his effort in attempting to explain the unusual magnetic spike at Colaba, India during the 1859 Carrington storm (Ohtani, 2022). The shape of the Colaba magnetic spike is certainly similar to the horizontal (*H*) geomagnetic field component profile observed during the Halloween storm at Tartu (MLAT = 54.6°) at 0650 UT on 29 October 2003. Ohtani (2022) points out that the cause of the *H* depression during the Halloween 29 October 2003 magnetic storm was "a dayside R1-field-aligned current (FAC) wedge" driven by dayside magnetic reconnection. The author indicates that the magnetic signature at Colaba during the Carrington storm was an FAC that was driven by interplanetary sheath fields (instead of a magnetic cloud (MC) as hypothesized by Tsurutani et al. (2003) and Lakhina et al. (2012)). Ohtani (2022) is in agreement with Siscoe et al. (2006) that the Carrington storm was caused by interplanetary sheath fields of 217 nT. We disagree with that possibility.

**Objections to Ohtani (2022)**

We have some severe difficulties with the Ohtani (2022) scenario. For the FAC to occur at Colaba (MLAT = 9°), the sheath interplanetary magnetic field (IMF) southward component ($B_s$) would have to be exceptionally large, far more intense than the IMF $B_s$ which caused the 29 October 2003 magnetic storm. It was shown by Tsurutani et al. (2018) that a normal ~5 cm$^{-3}$ solar wind density could explain the sudden impulse (SI$^+$) of ~120 nT observed at Colaba. From magnetohydrodynamics (MHD) theory, interplanetary shocks can compress the solar wind plasma densities and magnetic fields by a factor of the magnetosonic Mach number up to a factor of ~4 (Kennel et al., 1985). Since the typical solar wind IMF at 1 AU has a mean value of 5.6 nT with a standard deviation of 3.2 nT (Tsurutani et al., 2018), sheath magnetic field magnitudes between ~10 and 35 nT may be expected. These field values are too low to drive the magnetospheric current systems in to such extremely low latitudes, even if the magnetic field was directed totally southward.

Further, it is not clear which characteristics of R1-FAC wedge during the 2003 Halloween storm produced a magnetic profile at Tartu, and nowhere else, similar to the Colaba magnetic spike during the 1859 Carrington storm. The scenario of an FAC driven by an interplanetary sheath closing via a westward electrojet over Colaba seems difficult to imagine. Alex et al. (2006) have studied the Halloween storms from 29-31 October 2003 using interplanetary parameters and the magnetic data from the Alibag (ABG, MLAT = 9.7°) observatory, successor to the Colaba observatory, and also the equatorial station Tirunelveli (TIR, MLAT = -0.36°). The shock front associated with the fast interplanetary coronal mass ejection (ICME) impacted the Earth's magnetosphere at 0612 UT on 29 October 2003 and produced a SI$^+$ of +62 nT at ABG and a SI$^+$ of +113 nT at TIR at 0612 UT (cf. Figure 1 of Alex et al., 2006). In the initial part of the storm during 06-09 UT on 29 October 2003, which is the focus of study by Ohtani (2022), the highly fluctuating interplanetary $B_z$ produced fluctuations in the $H$ depression at ABG and TIR. The maximum dip in $H$ component at ABH was ~ -200 nT and at TIR, ~ -280 nT. Since the SI$^+$ as well as the $H$ depression at ABG are smaller than that at TIR, this cannot be reconciled with the R1-FAC wedge scenario which predicts a decreasing $H$ depression with decreasing magnetic latitudes.

**On high interplanetary sheath fields**

Ohtani (2022) has used the Siscoe et al. (2006) argument that for the Carrington storm the interplanetary sheath field was ~217 nT. We disagree with this hypothesis. For the studies of all available superstorms during the space age with intensities ~ -400 nT < Dst/SYM-H < -250 nT (e.g., Tsurutani et al., 1991; Echer et al., 2008; Meng et al., 2019) the IMF magnitudes associated with such storms had values between ~25 and ~75 nT. Of course the southward component of the field was only a fraction of the field magnitude. The above surveys did not include the 1989 Hydro-Quebec storm with peak intensity SYM-H = -710 nT/Dst = -589 nT because there was no available interplanetary data for that event. Extrapolation of the solar wind numbers from Tsurutani et al. (2003) gives an estimate of a Carrington MC magnetic field $B_z$ component of ~ -90 nT. The IMF magnitude would have to be even larger.

We would like to point out that Siscoe et al. (2006) based on their MHD simulation assumed an IMF strength of 5 nT and a fast shock compression of a factor of 4x giving an interplanetary sheath field of 20 nT, the same value as used in Tsurutani et al. (2018). We are in agreement of Siscoe et al. (2006) up to this point. Siscoe et al. realizing that they needed much larger magnetic field magnitudes for their interplanetary sheath scenario to work, invoked a Siscoe et al. (2002) MHD simulation study of the Earth's magnetosheath to obtain a maximum IMF strength of ~217 nT (a magnification factor of ~10). Based on other extrapolation models they obtained IMF values of 132 nT (Owens & Cargill, 2004) and a minimum of 66 nT. We do not believe an analogy to the Earth's magnetosheath is a correct one to use to estimate further magnification of interplanetary sheath magnetic fields. Such a magnification effect has not been observed in the Meng et al. (2019) study of superstorms, many of which were caused by interplanetary sheath magnetic fields. In general, magnetic field pileups at ICME boundaries have not been reported (or noticed by the authors) in interplanetary sheath studies.

Why does magnetic pileup occur in planetary magnetosheaths and not in interplanetary sheaths? Tsurutani et al. (1982: Figure 14) shows an example where the sheath magnetic field increases from ~5 nT just downstream of the shock to a maximum of ~10 nT at the Saturnian magnetopause, a factor of two times in amplification. The explanation of the physical cause of this effect has been called the "field line draping effect" (Midgley & Davis, 1963; Zwan & Wolf, 1976). Basically the mechanism is that the draped magnetic fields squeeze out the plasma along the lines of force leaving a low-β region behind (β is the ratio of the plasma thermal pressure to the magnetic pressure). So for an intial β = 1 plasma, typical of the solar wind, the magnetic field can intensify by a factor of ~2 if all the plasma has been ejected. Why doesn't the same effect happen at ICME sheaths? The scale size of an ICME is much, much larger than that of a planetary magnetosphere, perhaps ~0.3 AU or larger (Byrne et al., 2010). The plasma would take a long time to escape the vicinity of the ICME. Perhaps such an effect of high-sheath magnetic field magnitudes adjacent the antisunward side of ICMEs could be noticed by Voyager near the heliopause, but it does not seem to occur near 1 AU.

**On magnetic field profile of the Carrington storm**

Ohtani (2022) has questioned the rapid decay of the Colaba ~-1600 nT spike as possibly not being due to a decay of the Carrington storm time ring current. Siscoe et al. (2006) came up with a maximum H-depression of Dst = -850 nT. We disagree with these conclusions. We remind the reader that when Tsurutani et al. (2003) and Lakina et al. (2012) cited a value of Dst = -1760 nT for the Carrington storm, they were calculating the storm maximum intensity using the well-cited expression of Burton et al. (1975), not just an average Dst intensity. Prior to 1975 such high-speed ICMEs (1796 km s$^{-1}$ estimated by Tsurutani et al., 2003) were essentially unknown and the 1 min average SYM-H index was not available. The delay between the Colaba SI$^+$ and onset of the storm main phase was ~1 hour. For an 800 km s$^{-1}$ ICME, the delay would have been ~2+ hours, not so terribly unusual. We now use the much higher time resolution 1 minute average SYM-H index to identify storm maximum intensities. So, here in this paper we quote the SYM-H/Dst value of the Carrington storm as -1760 nT. It should be noted that in this paper we mention that the 1989 Hydro-Quebec storm is quoted as SYM-H = -710 nT, now slightly less than ½ the intensity of the Carrington event.

**Final Comments**

There is a way for sheath magnetic fields to attain very high values. This occurs if there are multiple shocks compressing a sheath around an ICME. An example of this was shown by Tsurutani et al. (2014) for the 7-17 March 2012 CAWSES II geomagnetic storm event. Three ICME shocks compressed the sheath field up to a ~68 nT peak value. However, this scenario does not work for the Carrington storm event. Each shock would compress the magnetosphere causing a SI$^+$. Such multiple SI$^+$s were not detected at Colaba during the 1-2 September 1859 storm.

A second way to get high sheath magnetic fields is via multiple MC-sheath interactions. Lakhina and Tsurutani (2016) have analyzed the SYM-H data for the 1989 Hydro-Quebec ~24 hour storm main phase. They have identified at least 2 and possibly 4 shock/sheath intervals and at least one MC interval contributing to the storm main phase of intensity SYM-H = -710 nT. What is interesting about this complex interplanetary event is that it had an aggregate speed that was quite slow and caused what we would call a "stealth magnetic storm", one whose onset and whose intensity would have been quite difficult to predict.

Wang et al. (2003) and Cerrato et al. (2011) discussed a complex interplanetary event that led to the magnetic storm (Dst = -387 nT) on 31 March 2001. For this event, three successive ICMEs were involved, two on 28 March with speeds of 427 km s$^{-1}$ and 519 km s$^{-1}$, and the third one on 29 March with speed of 942 km s$^{-1}$. The speeds are the projected CME speeds obtained from the SOHO/LASCO CME catalog (https://cdaw.gsfc.nasa.gov/CME_list/halo/halo.html). The high velocity of the last ICME is believed to have allowed it to overtake the former ones leading to compressed $B_z$ (northward) field ~55 nT in the sheath region. There is no indication that such an effect occurred for the Carrington event. For the Carrington storm, the SI$^+$ of 120 nT was followed about 1 hour later by the start of the main phase of the magnetic storm. An interpretation of these feature are that a fast ICME shock caused the SI$^+$, followed by a short-duration sheath and then southward $B_z$ fields in the MC caused the storm onset.

Tsurutani et al. (2018) have recently suggested a possible mechanism for the short duration of the Colaba spike. The rapid storm recovery may be caused by the storm time ring current particle losses due to electromagnetic ion cyclotron (EMIC) wave scattering associated with the much larger loss cone at low L. Tsurutani et al. (2018) also mentioned that other well known ring current loss processes such as charge exchange and Coulomb collisions (Kozyra et al., 1997; Jordanova et al., 1998) would be much more intense at low L during the Carrington storm as well. With the much stronger convection electric fields during the Carrington event, convection out the dayside magnetopause would also be more rapid.

It should be noted that the intensity of the Carrington magnetic storm (Dst/SYM-H peak = -1760 nT) was estimated by Tsurutani et al. (2003) and Lakhina et al. (2012) based on the location of the red (SAR/stable auroral red arc) aurora identified by Kimball (1960). Interplanetary parameters were studied to show that they were consistent with the SAR arc observations.

The work by Ohtani (2022) is interesting but we feel that it is not physically realistic.

**Acknowledgments**
GSL thanks the Indian National Science Academy, New Delhi for the support under the INSA-Honorary Scientist Scheme. The authors declare no real or perceived financial conflicts of interests. Work performed by RH was funded by the Science and Engineering Research Board (SERB Grant No. SB/S2/RJN-080/2018), a statutory body of the Department of Science and Technology (DST), Government of India through the Ramanujan Fellowship. We thank T. Hada and X. Meng for helpful scientific discussion.

**Open Research**

No data were analyzed in this article.